\documentclass[seceq]{ptptex}

\usepackage{graphicx}
\usepackage{wrapft}
\usepackage{amsfonts}
\usepackage{amssymb}
\usepackage{amsmath}
\usepackage{tabularx}
\newcommand{\dalm}{\kern1pt\vbox{\hrule height 0.9pt\hbox{\vrule width 0.9pt\hskip 2.5pt\vbox{\vskip 5.5pt}\hskip 3pt\vrule width 0.3pt}\hrule height 0.3pt}\kern1pt}

\notypesetlogo                       

\title{
Quantum Effect and Curvature Strength of Naked Singularities}

\author{
Umpei \textsc{Miyamoto},$^{1,}$\footnote{E-mail:umpei@gravity.phys.waseda.ac.jp} \ Hideki \textsc{Maeda}$^{2,}$\footnote{E-mail:hideki@gravity.phys.waseda.ac.jp} \ and Tomohiro \textsc{Harada}$^{3,}$\footnote{E-mail:T.Harada@qmul.ac.uk}%
}
\inst{
$^{1}$Department of Physics, Waseda University,
Tokyo 169-8555, Japan\\
$^{2}$Advanced Research Institute for Science and Engineering, Waseda University, Tokyo 169-8555, Japan\\
$^{3}$Astronomy Unit, School of Mathematical Sciences,
Queen Mary, University of London, Mile End Road, London E1 4NS, UK
}

\recdate{November 20, 2004}

\abst{
There are many solutions to the Einstein field equations that demonstrate 
naked singularity (NS) formation after regular evolution.
It is possible, however, that such a quantum effect as particle creation prevents
NSs from forming.
We investigate the relation between the curvature strength and
the quantum effects of NSs in a very wide 
class of spherical dust collapse.
Through a perturbative calculation,
we find that if the NS is very strong, the quantum particle
creation diverges as the Cauchy horizon is approached, while
if the NS is very weak, the creation should be finite.
In the context of cosmic censorship, strong NSs
will be subjected to the backreaction of 
quantum effects and may disappear or be hidden behind horizons, while weak NSs will not.
}

\begin{document}

\maketitle

\section{Introduction}
The cosmic censorship hypothesis (CCH) presents one of the most
important unsolved problems in general relativity.\cite{Penrose} \ Its
validity is often assumed in the analysis of physical phenomena in
strong gravitational fields. There are two versions of this
hypothesis. The weak hypothesis states that all singularities in
gravitational collapse are hidden within black holes. This version
implies the future predictability of the spacetime outside the event
horizon. The strong hypothesis asserts that no singularities visible to any
observer can exist. This version states that all physically reasonable
spacetimes are globally hyperbolic. Despite several attempts, neither
a general proof nor a precise mathematical formulation of the hypothesis has yet
been obtained. By contrast, some solutions of the Einstein field equation with regular initial conditions evolving into spacetimes containing naked singularities (NSs) have been found.\cite{Joshi,harada2004}

For a naked-singular spacetime to provide a counterexample of the CCH,
it is at least necessary that the Cauchy horizon (CH) be
stable. Although the CCH was originally stated in the classical context,
CHs may be unstable due to the backreaction of quantum effects, such as particle creation, i.e., particle creation may prevent NSs from forming. Study of such a possibility can be traced back to the pioneering works of Ford and Parker~\cite{FordParker} and Hiscock, Williams, and Eardley.\cite{HiscockWilliamsEardley} \ Ford and Parker considered the particle creation during the formation of a shell-crossing NS to obtain a finite amount of flux.\cite{FordParker} \ Hiscock \textit{et al.} considered the formation of a shell-focusing NS in the collapse of a null-dust fluid to obtain a diverging amount of flux.\cite{HiscockWilliamsEardley} \ Subsequently, such quantum phenomena have been studied in models of a self-similar dust,\cite{BarveSinghVazWitten,IguchiHarada,MiyamotoHarada} \ a self-similar null dust,\cite{SinghVaz,MiyamotoHarada} \ and an analytic dust \cite{HaradaIguchiNakao2000} models, for which the luminosities are found to diverge as negative powers of the time remaining to the CHs.
The analytic model describes the spherical dust collapse with an analytic initial density
profile with respect to
locally Cartesian coordinates. The analyticity of the initial density
profile and the self-similarity are incompatible in the spherically
symmetric dust model.
It is argued that the quantum radiation from a
strong NS, such as a shell-focusing one, must diverge as the CH is
approached,\cite{BarveSinghVazWitten,SinghVaz} \ although this has not yet been proved.
See Ref.~\citen{HaradaIguchiNakao2002} for a recent review of the quantum and classical radiation processes during NS formation.

Recently, two of the present authors computed the quantum radiation in
spherically symmetric self-similar collapse without specifying the
collapsing matter.\cite{MiyamotoHarada} \ It was found that in the
generic class of self-similar spacetimes resulting in a NS, the
luminosity of particle creation diverges as the inverse square of the
time remaining to the CH. Moreover, there was another interesting
result, leading us to the present study. 
In the self-similar collapse of a massless scalar field, described by the Roberts solution,\cite{Roberts} \ the luminosity remains finite at the CH. 
This result can be interpreted in terms of the curvature strength of the NS along the
CH. 
Although NSs forming in generic spherically symmetric self-similar
spacetimes are known~\cite{WaughLake} to satisfy
the strong curvature condition (SCC)~\cite{Tipler} along the CH,
 \ the NS appearing in the Roberts solution does not satisfy even the limiting focusing condition (LFC),~\cite{Krolak} which is weaker than the SCC.\footnote{
Following the work of Clarke and Kr\'{o}lak,\cite{ClarkeKrolak} \ consider a geodesic ($N$), affinely parameterized by $\kappa$, with tangent vector $k^{\mu}$, and terminating at or emanating from a singularity, where $\kappa=0$. If $\lim_{\kappa\to 0}\kappa^{2}R_{\mu\nu}k^{\mu}k^{\nu}\neq 0$ and $\lim_{\kappa\to 0}\kappa R_{\mu\nu}k^{\mu}k^{\nu}\neq 0$, where $R_{\mu\nu}$ is
the Ricci tensor, then the SCC and LFC are satisfied along $N$, respectively.
Since the quantity $R_{\mu\nu}k^{\mu}k^{\nu}$ for the Roberts solution indeed vanishes along the CH, the NS satisfies neither the SCC nor the LFC.}
The relation between the  curvature strength
and quantum effect of NSs has already been suggested
in Ref.~\citen{HaradaIguchiNakao2000}. 
It demonstrates the weak divergence of the
quantum radiation in the so-called analytic dust
model, in which the forming NS is known to be
weak.\cite{DeshingkarJoshiDwivedi,HaradaNakaoIguchi} 
However, a comprehensive
understanding of the relation between the curvature strength and the
quantum effects of NSs has not yet been realized.
The purpose here is to 
show how the amount of quantum radiation during the formation of NSs depends on such properties of singularities as the curvature strength. This analysis will help us obtain knowledge about the instability of CH, which should be predicted by a full semiclassical theory, taking into account the backreaction of quantum fields on gravity. In addition, it is shown here how the manner in which the quantized scalar fields are coupled to gravity changes the amount of quantum radiation. The dependence on the manner of coupling is important, because the CHs exhibit a semiclassical instability, caused by all fundamental quantum fields.

The organization of this paper is as follows. In \S~\ref{LTB}, we introduce a class of the Lema\^{\i}tre-Tolman-Bondi (LTB) solutions~\cite{LemaitreTolmanBondi} that result in the formation of a NS.
Then, the class of LTB solutions is divided into three sub-classes, depending on the curvature strength of the NSs in \S~\ref{strength}.
In \S~\ref{sec:local-map}, we calculate a map of null rays passing near
the NS. This plays crucial roles in estimating quantum emission in
the geometrical optics
approximation. In \S~\ref{Power}, the luminosity and total energy of emitted particles are computed. Section~\ref{Conclusion} is devoted to discussion.
Throughout this paper, the units for which $c=G=\hbar=1$ and the signature $(-,+,+,+)$ for spacetime metrics are used.
\section{Naked singularity in spherically symmetric dust collapse}
\subsection{Lema\^{\i}tre-Tolman-Bondi solutions admitting a naked singularity}
\label{LTB}
 The LTB solution,\cite{LemaitreTolmanBondi} \ which describes the
 collapse of a dust fluid, is written in a comoving coordinate system as
\begin{eqnarray}
ds^{2}&=&-dt^{2}+\frac{R^{\prime 2}}{1+f(r)}dr^{2}+R^{2}(t,r)d\Omega^{2},\nonumber \\
\dot{R}^{2}&=&\frac{F(r)}{R}+f(r),\label{Rdot}\\
\rho&=&\frac{F^{\prime}}{8\pi R^{2}R^{\prime}},\nonumber
\end{eqnarray}
where $\rho$ is the energy density, $d\Omega^{2}$ is the line element of a unit two-dimensional sphere, and the prime and dot denote the partial derivatives with respect to $r$ and $t$, respectively. Since we are concerned with the collapse of a dust fluid, we require $\dot{R}<0$. The arbitrary functions $F(r)$ and $1+f(r)>0$ are twice the conserved Misner-Sharp mass and the specific energy, respectively. In this paper, we only consider the case $f=0$, which is called `marginally bound collapse'.
In this case, Eq.~(\ref{Rdot}) is integrated to give
\begin{eqnarray}
R^{3}=\frac{9}{4}F(r)\left[t-t_{s}(r)\right]^{2},
\label{circumferential}
\end{eqnarray}
where $t_{s}(r)$ is an arbitrary function of $r$. The time $t=t_{s}(r)$
corresponds to the moment when a dust shell at $r$ meets the shell-focusing  singularity, which is defined by $R=0$. It is possible to choose $t_{s}(r)=r$ by scaling $r$.
LTB solutions can describe the formation of a shell-focusing NS from regular spacetimes. It has been shown that the shell-focusing singularity existing at $R=0$ with $r>0$ is completely spacelike,\cite{Christodoulou} \ and therefore our investigation will be confined to the singularity at $(t,r)=(0,0)$.

Here, we introduce a class of marginally bound LTB solutions in which the leading-order term of the mass function $F$ near the regular center takes the form
\begin{eqnarray}
F(r)=\frac{4\lambda^{3}}{9(\mu+1)^{3}}r^{3\mu+1}+\textrm{o}(r^{3\mu+1}),
\label{F}
\end{eqnarray}
where $\mu$ $(\geqslant 0)$ and $\lambda$ $(>0)$ are constants.\footnote{
This form of $F$ is chosen so that 
the initial density profile can be expanded in a power series in $R$ [see Eq.~(\ref{expanded-rho})] and the singularity
located at $(t, r)=(0, 0)$ is at least locally naked (see
Appendix~\ref{nakedness}). The nontrivial form of the factor and  power in
Eq.~(\ref{F}) are just for the simplicity of Eq.~(\ref{NGE}). Of course,
other choices of $F$ can allow the singularity to be naked.}
In Appendix~\ref{nakedness}, it is shown  that this class of LTB
spacetimes results in the formation of a shell-focusing NS, which could
be globally naked and therefore violate the weak version of CCH.
Radial null geodesics are described near the center by
\begin{eqnarray}
\frac{dt}{dr}=\pm R^{\prime}\simeq\pm\lambda r^{\mu}\mathcal{F}(t/r),\label{NGE}
\end{eqnarray}
where
\begin{eqnarray}
\mathcal{F}(z)\equiv\left[1-\frac{3\mu+1}{3(\mu+1)}z\right]\left(1-z\right)^{-1/3}.\nonumber
\end{eqnarray}
Here, the upper and lower signs correspond to outgoing and ingoing null geodesics, respectively.

We now consider the initial regular
density profile near the
regular center. The density profile at the initial time $t=t_{\rm in}<0$ is written,
\begin{eqnarray}
\rho(t_{\rm in},r)=\frac{1}{6\pi t_{\rm in}^{2}}\left[1+2\left(1+\frac{F}{rF^{\prime}}\right)\frac{r}{t_{\rm in}}+\textrm{O}\left((r/t_{\rm in})^{2}\right)\right]\nonumber.
\end{eqnarray}
Therefore, the initial density profile in terms of physical radius $R\propto r^{\mu+1/3}$ takes the form
\begin{eqnarray}
\rho_{\rm in}(R)=\rho_{0}+\rho_{1}R^{\gamma}+\cdots,
\label{expanded-rho}
\end{eqnarray}
where
\begin{eqnarray}
\gamma\equiv\frac{3}{3\mu+1},\;\;
\rho_{0}\equiv\frac{1}{6\pi t_{\rm in}^{2}},\;\;
\rho_{1}\equiv-\frac{(3\mu+2)(\mu+1)^{3/(3\mu+1)}}{3(3\mu+1)\pi\lambda^{3/(3\mu+1)}(-t_{\rm in})^{(9\mu+5)/(3\mu+1)}}.\label{coefficients}
\end{eqnarray}
The parameter $\gamma$ is in the region $0<\gamma\leqslant 3$ for $\mu\geqslant 0$. In particular, the analytic and self-similar LTB models are the cases with $\gamma=2$ ($\mu=1/6$) and $\gamma=3$ ($\mu=0$), respectively.
\subsection{Curvature strength of the naked singularities}
\label{strength}
The curvature strength of spacetime singularities is defined in the hope
that weak convergence would imply the extendibility of the spacetime in
a distributional sense. In this context, Tipler defined the
\textit{strong curvature condition} (SCC),\cite{Tipler} \ while
Kr\'{o}lak defined a weaker condition, which we call the
\textit{limiting focusing condition} (LFC).\cite{Krolak} \ The
sufficient and necessary conditions for the singularities in spherically
symmetric spacetimes with vanishing radial pressure satisfying the LFC
or SCC are given in simple forms.\cite{HaradaNakaoIguchi} \ 
Suppose a singularity is naked, and the relation between the circumferential radius $R$ and the Misner-Sharp mass $m$ is given by
\begin{eqnarray}
R\simeq 2y_{0}m^{\beta}
\label{R-m}
\end{eqnarray}
near the singularity along the null geodesics terminating at or emanating from the NS. The constant $\beta$ is in the region
\begin{eqnarray}
1/3<\beta\leqslant1.
\end{eqnarray}
The constants $y_{0}$ and $\beta$ are determined 
so that there is a positive finite limit
$y_{0}\equiv \lim_{m\to0}R/(2m^{\beta})$ along those
null geodesics. If
the geodesic satisfies the ``gravity-dominance
condition,''\cite{HaradaNakaoIguchi} 
the sufficient and necessary conditions are summarized as the following theorem.\cite{HaradaNakaoIguchi} \ \textit{For the radial null geodesic terminating from or emanating from the NS, if and only if $1/3<\beta<1/2$ is satisfied, neither the SCC nor the LFC holds, if and only if $1/2\leqslant\beta<1$ is satisfied, the LFC holds but the SCC does not, and if and only if $\beta=1$, both the SCC and the LFC hold.}

For the LTB spacetimes we consider, from
Eq.~(\ref{nullray}), we find that $R\propto r^{\mu+1}$ holds along the outgoing null
geodesic emanating from the NS, i.e., along the CH.\footnote{The null
ray given by Eq.~(\ref{nullray}) is an asymptotic solution of
the null geodesic equation with the boundary condition $t(0)=0$. In fact, Eq.~(\ref{nullray}) is the
\textit{earliest} null ray that 
emanates from the singularity.
This was proved by Christodoulou,~\cite{Christodoulou}
and that proof can be easily generalized to other cases.}
Also, we find that $m=F(r)/2\propto r^{3\mu+1}$ holds along the outgoing null geodesic near the singularity. Therefore, the constant $\beta$ is obtained as
\begin{eqnarray}
\beta=\frac{\mu+1}{3\mu+1}=\frac{2\gamma+3}{9}.
\label{beta}
\end{eqnarray}
With the above theorem, proved by Harada \textit{et al.}, the NS for $0<\gamma<3/4$ satisfies neither the LFC nor the SCC along the outgoing null geodesic emanating from the NS. The NS for $3/4\leqslant\gamma<3$ does not satisfy the SCC, but it does satisfy the LFC. The NS for $\gamma=3$ satisfies both the LFC and the SCC. (See also Table~\ref{tb:minimal} for the relation between $\gamma$ and the curvature strength of NSs.)

\section{Map of null rays passing near the naked singularity}
\label{sec:local-map}
\subsection{Local map}
\begin{figure}[bhtp]
\begin{center}
\includegraphics[width=7cm]{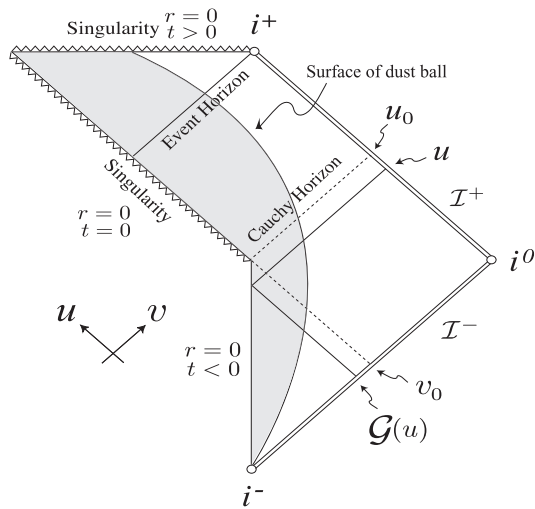}
\caption{\label{fg:conformal}
A possible causal structure of the LTB spacetimes considered in this article. A null singularity occurs at the spacetime point $(u_{0},v_{0})$, and it is visible from $\mathcal{I}^{+}$, where $(u,v)$ is a suitable double null coordinate system. An outgoing null ray $u=[\mathrm{constant}]$ can be traced backward in time from $\mathcal{I}^{+}$ to $\mathcal{I}^{-}$. It turns out to be an ingoing null ray $v=\mathcal{G}(u)$ after passing the regular center, located at $r=0$, with $t<0$. The outgoing null ray $u=u_{0}$ and the ingoing null ray $v=v_{0}$ represent the CH and the null ray that terminates at the NS, respectively.
}
\end{center}
\end{figure}

The \textit{global map} $v=\mathcal{G}(u)$ is defined as the relation
between the moments when a null ray leaves $\mathcal{I}^{-}$ and when it
terminates at $\mathcal{I}^{+}$ after passing through the regular center
(see Fig.~\ref{fg:conformal}). This global map plays crucial roles in
estimating the quantum radiation with the geometrical optics approximation. The
global map cannot be obtained without solving the null geodesic equation
from $\mathcal{I}^{-}$ to $\mathcal{I}^{+}$ globally. 

\begin{figure}[bhtp]
\begin{center}
\includegraphics[width=7cm]{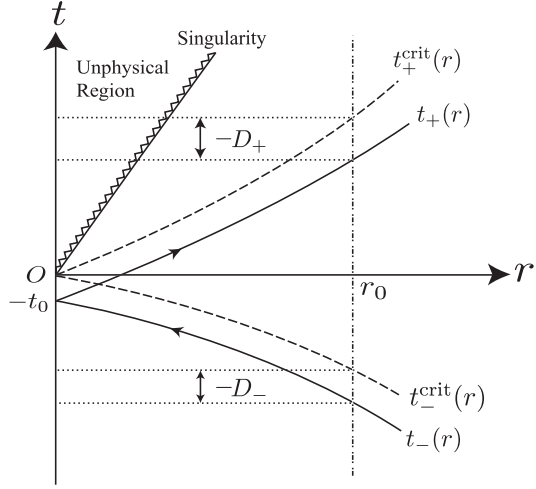}
\caption{\label{fg:map}
A schematic spacetime diagram of the 
region filled with dust, with the illustration of the local map defined in \S~\ref{sec:local-map}. A pair consisting of an ingoing null ray $t_{-}(r)$ and an outgoing null ray $t_{+}(r)$ is depicted (solid curve). It passes near the NS, located at $(t, r)=(0, 0)$ . Null rays terminating at and emanating from the NS, $t^{\rm crit}_{-}(r)$ and $t^{\rm crit}_{+}(r)$, are also depicted (dashed curves). The latter is the CH. A comoving observer is located at $r=r_{0}=[\textrm{constant}]$. The local map is defined as the relation between $t_{-}(r_{0})$ and $t_{+}(r_{0})$.
}
\end{center}
\end{figure}

However, the main
properties of the global map should be determined by the
behavior of null rays near the singularities, since the particle creation
should be caused by the curvature around singularities. From this point of
view, Tanaka and Singh proposed an alternative map, which we call the
\textit{local map}.\cite{TanakaSingh} \ They considered an observer on a
comoving shell who sends ingoing null rays. These null rays are
reflected at the regular center, and then return to the same comoving
observer. A radial null geodesic crosses a comoving shell located at a
fixed comoving radius $r$ before and after the reflection at the
center. Thus these null rays define a map between the sending time and
the receiving time measured by the proper time for the comoving
observer. (See Fig.~\ref{fg:map} for a schematic illustration of the
local map.) 

The local map is conjectured to have the same structure as the
global map, because there are no singular features in the map between the
proper time on a comoving shell at a finite distance and those measured
by the null coordinates naturally defined at infinity.
This conjecture has been confirmed for the self-similar
dust model~\cite{TanakaSingh,MiyamotoHarada} and the analytic
dust model,\cite{TanakaSingh}\ which is a non-self-similar spacetime. Two of the present authors generalized
the local map to general self-similar spacetimes, and the validity of such a local map was
confirmed for the self-similar Vaidya
model.\cite{MiyamotoHarada} \ Therefore, one can 
assume that the
local map has the same structure as the global map in the LTB spacetimes. 

In this section, we calculate the local map by solving the radial null
geodesic equation for the non-self-similar LTB spacetimes ($\mu>0$).
The local map for the self-similar LTB spacetime ($\mu=0$) was 
obtained in Refs.~\citen{TanakaSingh} and \citen{MiyamotoHarada}. 

\subsection{Outline of the calculation of local map}
\label{sec:outline}
To make the notation simple, we replace the coordinates $r$ and $t$ with $r^{\prime}$ and $t^{\prime}$ given by
\begin{eqnarray*}
r'&=&\lambda^{1/\mu}r,\\
t'&=&\lambda^{1/\mu}t,
\end{eqnarray*}
and we write $r'$ and $t'$ as $r$ and $t$ hereafter in this section.
Then, the null geodesic equation~(\ref{NGE}) becomes
\begin{eqnarray}
\frac{dt}{dr}=\pm r^{\mu}\mathcal{F}(t/r).
\label{NGE2}
\end{eqnarray}

Before beginning the calculation, it is 
convenient to summarize the calculational scheme for
obtaining the local map that we implement in the following subsections.
To obtain the local map, we have to obtain solutions
of Eq.~(\ref{NGE2}) near $r=0$, which correspond
to radial null rays passing through
the center at $t=t(0)=-t_{0}$ ($t_{0}>0$),
and then take the limit $t_{0}\to 0$. 
We cannot expect, however, to have 
general exact solutions to this equation. 
Hence, here we adopt the following scheme to obtain the local map.
First, we apply three different approximation
regimes, A, B and C, and find three kinds of 
expressions, $t=t^{\rm A}(r)$, 
$t=t^{\rm B}(r)$ and $t=t^{\rm C}(r)$, respectively. 
Next, we show that these three regimes have an overlapping 
region, where all three approximations are valid
and the obtained approximate solutions can be matched
with each other.  
Finally, we calculate the local map, that is, we calculate the 
relation between the sending time and the receiving time of the 
null ray at a comoving observer near the center.

\begin{figure}[bhtp]
\begin{center}
\includegraphics[width=7cm]{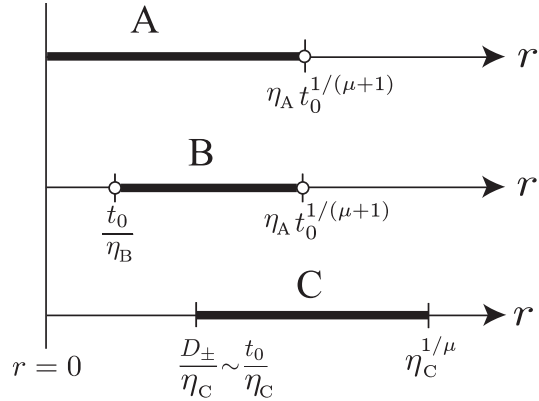}
\caption{\label{fg:regime}
A schematic illustration of the regions in which the regimes A, B and C are valid.
The regions corresponding to the regimes A, B and C are $0\leqslant r < \eta_{\rm A} t_{0}^{1/(1+\mu)}$, $t_{0}/\eta_{\rm B}< r < \eta_{\rm A} t_{0}^{1/(1+\mu)}$ and $D_{\pm}/\eta_{\rm C}\sim t_{0}/\eta_{\rm C} \lesssim r \lesssim \eta_{\rm C}^{1/\mu}$, respectively, where $\eta_{\rm X}$ $\ll 1$ ($\rm{X}=\rm{A}, \rm{B}, \rm{C}$) is a constant independent of $t_{0}$. The region for B is included in that for A. 
The regions for B and C exist if
 $t_{0}<(\eta_{\rm A}\eta_{\rm B})^{(\mu+1)/\mu}$ and $t_{0}< \eta_{\rm C}^{(\mu+1)/\mu}$ are satisfied, respectively. The regions for B and C overlap if $t_{0}<(\eta_{\rm A}\eta_{\rm C})^{(\mu+1)/\mu}$. All these conditions are satisfied 
for $t_{0}<\min[(\eta_{\rm A}\eta_{\rm B})^{(\mu+1)/\mu},\;(\eta_{\rm A}\eta_{\rm C})^{(\mu+1)/\mu}]$. In the limit $t_{0}\to 0$, the regions for A and B shrink to zero, but the region for C, where the comoving observer should be located, remains finite.
}
\end{center}
\end{figure}

In \S~\ref{sec:near}, we consider approximation regime A,
in which $0\leqslant r <\eta_{\rm A} t_{0}^{1/(1+\mu)}$ is satisfied. Here, $\eta_{\rm A}$
($\ll 1$) is a positive constant, independent of $t_{0}$.
In this regime, we can treat the center, and then we can relate 
the ingoing and outgoing null rays, which
reach the center at the same time, $t=-t_{0}$.
In \S~\ref{sec:nearB}, we consider approximation regime B,
in which $t_{0}/\eta_{\rm B}<r<\eta_{\rm A} t_{0}^{1/(1+\mu)}$ is satisfied.
Here, $\eta_{\rm B}$ ($\ll 1$) is a positive constant.
This regime can exist only when $t_{0}<\left(\eta_{\rm A}\eta_{\rm B}\right)^{(1+\mu)/\mu} $. 
Although regime B is completely included in regime A,
regime B enables us to have an explicit expression for 
solutions, and it is therefore essential to obtain the local map.
In \S~\ref{sec:far}, we consider approximation regime C,
where $t/r\ll 1$ is assumed. 
When we choose $t/r$ to be $\textrm{O}(\eta_{\rm C})$, where $\eta_{\rm C}$ ($\ll 1$) is a sufficiently small constant, 
it turns out that the approximation is valid for 
$t_{0}/\eta_{\rm C} \lesssim r \lesssim \eta_{\rm C}^{1/\mu}$.
Regimes A and B obviously have an overlapping region.
When we take the limit $t_{0}\to 0$, regime B and regime C also
have an overlapping region, and regime C is valid 
at finite radius. (See also Fig.~\ref{fg:regime} for the illustration of regions,
where each regime is valid.)
In \S~\ref{sec:matching}, we implement the matching 
between the approximate solutions $t=t^{\rm B}(r)$ and $t=t^{\rm C}(r)$
in the overlapping region.
In \S~\ref{sec:local_map_calc},
the local map is finally obtained for a comoving observer 
at a finite radius in the region for regime C.   
This is the generalization of the result of Tanaka and 
Singh.\cite{TanakaSingh} 
\subsection{Regime A: $0\leqslant r <\eta_{\rm A} t_{0}^{1/(1+\mu)}$}
\label{sec:near}
To find a null geodesic for $0\leqslant r <\eta_{\rm A} t_{0}^{1/(1+\mu)}$, 
we set
\begin{eqnarray}
\frac{r}{t_{0}^{1/(\mu+1)}}&=&\epsilon \zeta, \label{epsilon}\\
\frac{r}{t_{0}}&=&\frac{\zeta}{\delta},\label{delta}
\end{eqnarray}
where $\epsilon$ and $\delta$ are constants, and $\zeta$ is variable for $r$.
Here, we assume that $\epsilon<\eta_{\rm A}$
and $\zeta$ is $\textrm{O}(1)$.
The quantity $\delta $ is introduced for later convenience and is not necessarily small in regime A.
From Eqs.~(\ref{epsilon}) and (\ref{delta}), the following relations hold:
\begin{eqnarray}
t_{0}&=&\epsilon^{(\mu+1)/\mu}\delta^{(\mu+1)/\mu},\label{t02}\\
r&=&\epsilon^{(\mu+1)/\mu}\delta^{1/\mu}\zeta\label{rrho}.
\end{eqnarray}
The null geodesic $t=t^{\rm A}(r)$ can be expanded in $\epsilon$ as follows:
\begin{eqnarray}
t^{\rm A}(r)=-t_{0}+\sum_{n=1}^{\infty}\epsilon^{(\mu+1)(n\mu+1)/\mu}t^{\rm A}_{n}(\zeta),
\label{expansion1}
\end{eqnarray}
where $t^{\rm A}_{n}(\zeta)$ ($n=1,2,\cdots$)
are functions of $\zeta$ of order unity. 
Substituting Eq.~(\ref{expansion1}) into Eq.~(\ref{NGE2}), we obtain the 
differential equations
\begin{eqnarray}
\frac{dt^{\rm A}_{1}}{d\zeta}(\zeta)&=&\pm\delta^{(\mu+1)/\mu}\zeta^{\mu}\mathcal{F}(-\delta/\zeta),
\label{dt1drho}\\
\frac{dt^{\rm A}_{2}}{d\zeta}(\zeta)&=&\pm\delta\zeta^{\mu-1}\mathcal{F}^{\prime}(-\delta/\zeta)t^{\rm A}_{1}(\zeta),\label{dt2drho}\\
\frac{dt^{\rm A}_{3}}{d\zeta}(\zeta)&=&\pm\frac{1}{2}\delta^{(\mu-1)/\mu}\zeta^{\mu-2}\mathcal{F}^{\prime\prime}(-\delta/\zeta)(t^{\rm A}_{1})^{2}(\zeta)\pm\delta\zeta^{\mu-1} \mathcal{F}^{\prime}(-\delta/\zeta)t^{\rm A}_{2}(\zeta)\label{dt3drho},
\end{eqnarray}
and so on, where $\mathcal{F}^{\prime}$ and $\mathcal{F}^{\prime\prime}$ denote derivatives of $\mathcal{F}$ with respect to its argument.
Since we have $t^{\rm A}(0)=-t_{0}$, $t^{\rm A}_{n}(0)=0$ must be satisfied 
for $n\geqslant 1$.
Equation~(\ref{dt1drho}) can be integrated immediately to give
\begin{eqnarray}
t^{\rm A}_{1}(\zeta)
=\pm\frac{1}{\mu+1}\delta^{(\mu+1)/\mu}\zeta^{\mu+1}(1+\delta/\zeta)^{2/3}.
\label{t1}
\end{eqnarray}
With $t^{\rm A}_{1}(\zeta)$ obtained above, 
Eq.~(\ref{dt2drho}) is integrated to give
\begin{eqnarray}
t^{\rm A}_{2}(\zeta)&=&\frac{1}{\mu+1}\delta^{(2\mu+1)/\mu}\int^{\zeta}_{0}\hat{\zeta}^{2\mu}\mathcal{F}^{\prime}(-\delta/\hat{\zeta})(1+\delta/\hat{\zeta})^{2/3}d\hat{\zeta}\\
&=&\delta^{(\mu+1)(2\mu+1)/\mu}I_{2}(\zeta/\delta),
\end{eqnarray}
where
\begin{eqnarray}
I_{2}(y)\equiv\frac{1}{\mu+1}\int^{y}_{0}x^{2\mu}\mathcal{F}^{\prime}(-1/x)(1+1/x)^{2/3}dx.
\label{I2}
\end{eqnarray}
This integration cannot be expressed in terms of elementary functions. 
In a similar way, one can write $t^{\rm A}_{3}(\zeta)$ in the integral form
\begin{eqnarray}
t^{\rm A}_{3}(\zeta)&=&\pm\int^{\zeta}_{0}\biggl[\frac{1}{2(\mu+1)^{2}}\delta^{(3\mu+1)/\mu}\hat{\zeta}^{3\mu}\mathcal{F}^{\prime\prime}(-\delta/\hat{\zeta})(1+\delta/\hat{\zeta})^{4/3}\nonumber\\
&&\hspace{4cm}+\delta^{(2\mu^{2}+4\mu+1)/\mu}\hat{\zeta}^{\mu-1}\mathcal{F}^{\prime}(-\delta/\hat{\zeta})I_{2}(\hat{\zeta}/\delta)\biggr]d\hat{\zeta}\nonumber\\
&=&\pm\delta^{(\mu+1)(3\mu+1)/\mu}I_{3}(\zeta/\delta),\nonumber
\end{eqnarray}
where
\begin{equation}
I_{3}(y)\equiv\int^{y}_{0}\left[\frac{1}{2(\mu+1)^{2}}x^{3\mu}\mathcal{F}^{\prime\prime}(-1/x)(1+1/x)^{4/3}+x^{\mu-1}\mathcal{F}^{\prime}(-1/x)I_{2}(x)\right]dx.
\end{equation}
It should be noted that we can safely take the limit $r\to 0$
in this regime, because we do not assume that $\delta$ is small.
\subsection{Regime B: $t_{0}/\eta_{\rm B}<r<\eta_{\rm A} t_{0}^{1/(1+\mu)}$}
\label{sec:nearB}
For the approximation regime B, we additionally 
assume that $\delta<\eta_{\rm B}$.
This also requires that $t_{0}\ll 1$, from Eq.~(\ref{t02}).
The approximate solution $t=t^{\rm B}(r)$ can be obtained
by making approximation $t=t^{\rm A}(r)$ with $\delta \ll 1$.
Hence, we define $t^{\rm B}_{n}(\zeta)$ 
as the function which is obtained by approximating 
$t^{\rm A}_{n}(\zeta)$ with $\delta\ll 1$. 
Thus, we have 
\begin{equation}
t^{\rm B}_{1}=\pm \frac{1}{\mu+1}\delta^{(\mu+1)/\mu}\zeta^{\mu+1}
\left[1+\frac{2}{3}\zeta^{-1}\delta +\textrm{O}(\delta^{2})\right].
\label{t1B}
\end{equation}
The approximate form of $t^{\rm A}_{2}$ for $\delta \ll 1$ can be obtained 
by using the asymptotic form of $I_{2}(y)$ for large $y$ in
Eq.~(\ref{I2}) given by
\begin{eqnarray}
I_{2}(y)=C_{2}-\frac{2\mu}{3(\mu+1)^{2}(2\mu+1)}y^{2\mu+1}\left[1+\textrm{O}(1/y)\right],\nonumber
\end{eqnarray}
where $C_{2}$ is a constant. Except for the term $C_{2}$, all terms are completely determined by integrating the expanded integrand. 
Therefore, $t^{\rm B}_{2}(\zeta)$ is obtained as
\begin{eqnarray}
t^{\rm B}_{2}(\zeta)=C_{2}\delta^{(\mu+1)(2\mu+1)/\mu}-\frac{2\mu}{3(\mu+1)^{2}(2\mu+1)}\delta^{(2\mu+1)/\mu}\zeta^{2\mu+1}\left[1+\textrm{O}(\delta)\right].
\label{t2}
\end{eqnarray}
Similarly, the asymptotic form of $I_{3}$ is given by
\begin{equation}
I_{3}(y)=C_{3}+\left[\frac{2\mu^{2}+\mu+1}{9(\mu+1)^{3}(2\mu+1)(3\mu+1)}y^{3\mu+1}-\frac{2C_{2}}{3(\mu+1)}y^{\mu}\right]\left[1+\textrm{O}(1/y)\right],\nonumber
\end{equation}
and hence $t^{\rm B}_{3}(\zeta)$ is 
\begin{multline}
\label{t3}
t^{\rm B}_{3}(\zeta)=\pm C_{3}\delta^{(\mu+1)(3\mu+1)/\mu}\pm
\frac{2\mu^{2}+\mu+1}{9(\mu+1)^{3}(2\mu+1)(3\mu+1)}
\delta^{(3\mu+1)/\mu}\zeta^{3\mu+1}\left[1+\textrm{O}(\delta)\right] \\
\mp\frac{2C_{2}}{3(\mu+1)}\delta^{(2\mu^{2}+4\mu+1)/\mu}\zeta^{\mu}
\left[1+\textrm{O}(\delta)\right].
\end{multline}
Then, an approximate solution is obtained 
from Eqs.~(\ref{t02})-(\ref{expansion1}) and (\ref{t1B})-(\ref{t3}) as
\begin{multline}
\label{tB}
t^{\rm B}(r)=-t_{0}+C_{2}t_{0}^{2\mu+1}\pm C_{3}t_{0}^{3\mu+1}\\
+\left[\pm\frac{1}{\mu+1}r^{\mu+1}-\frac{2\mu}{3(\mu+1)^{2}(2\mu+1)}r^{2\mu+1}\pm\frac{2\mu^{2}+\mu+1}{9(\mu+1)^{3}(2\mu+1)(3\mu+1)}r^{3\mu+1}\right]\\
\times\left[1+\textrm{O}(\delta)\right].
\end{multline}
For later use, we have obtained explicit expression up 
to third order. It is straightforward to compute 
higher orders. It should be noted that we cannot take the 
limit $r\to 0$ in ~(\ref{tB}).
\subsection{Regime C: $t/r\ll 1$}
\label{sec:far}
Suppose $t/r\ll 1$. Here, we introduce the condition $\eta_{\rm C}\ll 1$, which controls 
the order of $t/r$; specifically $t/r$ is $\textrm{O}(\eta_{\rm C})$. In this approximation regime,
we can expand $f$ in $t/r$
on the right-hand side of Eq.~(\ref{NGE2}) and obtain
the expanded form of solutions. 

Let us consider the critical outgoing and 
ingoing null geodesics $t=t^{\rm crit}_{\pm}(r)$,
which emanate from and terminate at the NS, i.e., 
$t=r=0$.
Then $t=t^{\rm crit}_{+}(r)$ gives the CH, by definition.
If we assume that $t/r$ is $\textrm{O}(\eta_{\rm C})$, these geodesics are obtained by expanding the 
null geodesic equation~(\ref{NGE2}) in power series in $r$ with the boundary condition $t(0)=0$:
\begin{multline}
\label{CHs}
t^{\rm crit}_{\pm}(r) = \pm\frac{1}{\mu+1}r^{\mu+1}
-\frac{2\mu}{3(\mu+1)^{2}(2\mu+1)}r^{2\mu+1} \\
\pm\frac{2\mu^{2}+\mu+1}{9(\mu+1)^{3}(2\mu+1)(3\mu+1)}r^{3\mu+1}
+\textrm{O}(\eta_{\rm C}^{4}).
\end{multline}
Here, the upper (lower) sign corresponds to an emanating (terminating) null
ray. For $0< r\lesssim \eta_{\rm C}^{1/\mu}$,  $t/r$ is $\textrm{O}(\eta_{\rm C})$ on the critical null rays, and this approximation is valid.
That is, the approximation regime C is valid for $0< r \lesssim \eta_{\rm C}^{1/\mu}$ on the critical null rays.

However, we are interested in null rays that correspond to times slightly before the times of these critical null rays. 
We expand the solution as
\begin{eqnarray}
t=t^{\rm C}(r)=\sum_{n=1}^{\infty} t^{\rm C}_{n}(r),
\label{expansion2}
\end{eqnarray}
where we assume that $t^{\rm C}_{n}(r)/r$ is $\textrm{O}(\eta_{\rm C}^{n})$.
Substituting Eq.~(\ref{expansion2}) into Eq.~(\ref{NGE2})
and expanding the right-hand side, 
the following differential equations are obtained:
\begin{eqnarray}
\frac{d t^{\rm C}_{1}}{d r }&=&\pm r^{\mu},\label{dtau1dsigma}\\
\frac{d t^{\rm C}_{2}}{d r }&=&\mp\frac{2\mu}{3(\mu+1)}r ^{\mu-1} t^{\rm C}_{1}(r),\label{dtau2dsigma}\\
\frac{d t^{\rm C}_{3}}{d r }&=&\mp\frac{2\mu}{3(\mu+1)}r^{\mu-1} t^{\rm C}_{2}( r )\mp\frac{\mu-1}{9(\mu+1)}r^{\mu-2} (t^{\rm C}_{1})^{2}( r ).\label{dtau3dsigma}
\end{eqnarray}
These equations are integrated to yield
\begin{eqnarray}
 t^{\rm C}_{1}( r )&=&D_{\pm}\pm\frac{1}{\mu+1} r ^{\mu+1},\label{tau1}\\
 t^{\rm C}_{2}( r )&=&\mp\frac{2}{3(\mu+1)}D_{\pm} r ^{\mu}-\frac{2\mu}{3(\mu+1)^{2}(2\mu+1)} r ^{2\mu+1},\label{tau2}\\
 t^{\rm C}_{3}( r )&=&\mp\frac{1}{9(\mu+1)}D_{\pm}^{2} r ^{\mu-1}+\frac{1}{9\mu(\mu+1)}D_{\pm} r ^{2\mu}\nonumber\\
&&\hspace{4cm}\pm\frac{2\mu^{2}+\mu+1}{9(\mu+1)^{3}(2\mu+1)(3\mu+1)} r ^{3\mu+1},\label{tau3}
\end{eqnarray}
where $D_{\pm}$ is an integration constant that comes from the
integration of Eq.~(\ref{dtau1dsigma}). The integration constants for
Eqs.~(\ref{dtau2dsigma}) and (\ref{dtau3dsigma}) are set to zero. From Eqs.~(\ref{expansion2}) and (\ref{tau1})-(\ref{tau3}), the
solution takes the following form:
\begin{eqnarray}
\label{tau}
t^{\rm C}(r)&=&D_{\pm} \nonumber \\
&\pm&\frac{1}{\mu+1}r^{\mu+1}\nonumber\\
&\mp&\frac{2}{3(\mu+1)}D_{\pm}r^{\mu}
-\frac{2\mu}{3(\mu+1)^{2}(2\mu+1)}r^{2\mu+1} \nonumber \\
&\mp&\frac{1}{9(\mu+1)}D_{\pm}^{2}r^{\mu-1}
+\frac{1}{9\mu(\mu+1)}D_{\pm}r^{2\mu}
\pm\frac{2\mu^{2}+\mu+1}{9(\mu+1)^{3}(2\mu+1)(3\mu+1)}r^{3\mu+1}\nonumber \\
&+&\textrm{O}(\eta_{\rm C}^{4}).
\end{eqnarray}

Now, we can examine in what region the solution (\ref{tau}) is valid.
In order for the expansion (\ref{expansion2}) to be valid, $t^{\rm C}_{n}(r)/r$ must be $\textrm{O}(\eta_{\rm C}^{n})$.
This is the case if
\begin{equation}
\max\left(r^{\mu},\frac{D_{\pm}}{r}\right)=\textrm{O}(\eta_{\rm C}).\nonumber
\end{equation}
If $D_{\pm}$ is sufficiently small, the above condition 
implies
\begin{equation}
D_{\pm}/\eta_{\rm C} \lesssim r \lesssim \eta_{\rm C}^{1/\mu}.\nonumber
\end{equation}
This is the region in which approximation regime C is valid for the null geodesic
passing through the center at $t(0)=-t_{0}<0$. The relation between $D_{\pm}$ and $t_{0}$ is
elucidated in the following subsection.
\subsection{Matching the approximation regimes}
\label{sec:matching}
Since we are interested in the null rays that are close to 
the critical null geodesics, we take the limit $t_{0}\to 0$.
The region for approximation regime B
is completely included in that for regime A, and
the matching is trivially implemented.
If we assume $D_{\pm}=\textrm{O}(t_{0})$, 
the region in which regime C is valid must have 
an overlap with the region in which regime B is valid
if $t_{0}\lesssim (\eta_{\rm A}\eta_{\rm C})^{(\mu+1)/\mu}$.
This means that we can relate the integration constants
$D_{\pm}$, which appear in the regime C solution $t=t^{\rm C}(r)$,
to $t_{0}$, which appears in the regime B solution $t=t^{\rm B}(r)$,
by matching these two solutions in the overlapping region.  

The different expressions $t=t^{\rm B}(r)$
and $t=t^{\rm C}(r)$ for the null geodesic 
are obtained independently in
\S\S~\ref{sec:nearB} and \ref{sec:far}. 
We find that the solution $t=t^{\rm B}(r)$ given by Eq.~(\ref{tB})
coincides with the solution $t=t^{\rm C}(r)$ given by Eq.~(\ref{tau})
at several lowest orders if the constant terms 
satisfy the following relation:
\begin{eqnarray}
D_{\pm}\simeq -t_{0}+C_{2}t_{0}^{2\mu+1}\pm C_{3}t_{0}^{3\mu+1}.
\label{D-t0}
\end{eqnarray}
This verifies the validity of the assumption that 
$D_{\pm}$ is sufficiently small and $D_{\pm}$ is $\textrm{O}(t_{0})$
in the limit $t_{0}\to 0$.
This also implies that the approximation regime C is valid
for $t_{0}/\eta_{\rm C}\lesssim r\lesssim \eta_{\rm C}^{1/\mu}$.
Here, let us see the condition on $t_{0}$ which must be satisfied for the matching.
$t_{0}$~has to  satisfy
\begin{eqnarray}
t_{0}<\min\left((\eta_{\rm A}\eta_{\rm B})^{(\mu+1)/\mu},\;(\eta_{\rm A}\eta_{\rm C})^{(\mu+1)/\mu}\right),
\end{eqnarray}
so that the region for the regime B and the overlapping region for the regimes B and C can exist.
It is possible to take such a small $t_{0}$, because we are interested in the limit $t_{0}\to 0$. (See also Fig.~\ref{fg:regime}.)
\subsection{Obtaining the local map}
\label{sec:local_map_calc}
To obtain the local map for this spacetime,
we consider a comoving observer at $r=r_{0}$, where $r_{0}$ satisfies
$r_{0}\lesssim \eta_{\rm C}^{1/\mu}$. 
As time proceeds, this observer approaches the ingoing critical
null ray $t=t^{\rm crit}_{-}(r)$ and therefore enters the region
$t_{0}/\eta_{\rm C}\lesssim r \lesssim \eta_{\rm C}^{1/\mu}$, where approximation 
regime C is valid.
Then, $t_{\pm}(r_{0})=t^{\rm C}_{\pm}(r_{0})$ are regarded as 
the sending time and receiving time of the null geodesic,
where the signs $\pm$ are introduced to distinguish
the outgoing and ingoing null rays. 
From Eq.~(\ref{tau}), the constant $D_{\pm}$ is specified as
\begin{eqnarray}
-D_{\pm}= t^{\rm crit}_{\pm}(r_{0})- t_{\pm}(r_{0})+\textrm{O}(t_{0}\eta_{\rm C}),
\nonumber
\end{eqnarray}
where $t^{\rm crit}_{\pm}(r_{0})$ is the moment when the observer crosses 
the null geodesic~(\ref{CHs}).
Therefore, if we take $\eta_{\rm C}$ to be sufficiently small,
$D_{-}$ ($D_{+}$) can be interpreted as the time 
difference between the moments when the observer emits (receives) 
the null ray and crosses the null ray terminating at (emanating from) the NS. (See also Fig.~\ref{fg:map}.) 

Since we consider a set consisting of an ingoing null ray 
which reaches the center at $t=-t_{0}$
and an outgoing null ray which can be regarded
as a reflected ray of the former at $t=-t_{0}$,
we pick up both the ingoing and outgoing null rays with 
the same value for $t_{0}$.
When we eliminate $t_{0}$ from Eq.~(\ref{D-t0}) for both signs,
we have the relation between $D_{+}$ and $D_{-}$
\begin{eqnarray}
D_{-}\simeq D_{+}-2C_{3}(-D_{+})^{3\mu+1},\nonumber
\end{eqnarray}
which can be rewritten in terms of the sending and receiving 
times $ t_{\pm}(r_{0})$ at $r=r_{0}$ for sufficiently small $r_{0}$
as
\begin{eqnarray}
 t^{\rm C}_{-}(r_{0})\simeq t^{\rm crit}_{-}(r_{0})-\left[t^{\rm crit}_{+}(r_{0})- 
t^{\rm C}_{+}(r_{0})\right]-2C_{3}\left[ t^{\rm crit}_{+}(r_{0})- 
t^{\rm C}_{+}(r_{0})\right]^{3\mu+1}.
\label{eq:local-map_}
\end{eqnarray}
Equation~(\ref{eq:local-map_}) is the very local map, relating the
moments when the comoving observer locating at $r=r_{0}$ sends the
ingoing null ray and receives the reflected outgoing null rays. If 
we revive $\lambda$, the final result becomes
\begin{eqnarray}
 t^{\rm C}_{-}(r_{0})\simeq t^{\rm crit}_{-}(r_{0})-\left[ t^{\rm crit}_{+}(r_{0})- 
t^{\rm C}_{+}(r_{0})\right]-2C_{3}\lambda^{3}\left[ t^{\rm crit}_{+}(r_{0})- 
t^{\rm C}_{+}(r_{0})\right]^{3\mu+1}.
\label{eq:local-map}
\end{eqnarray}
Now, note that the comoving observer must be in the region 
$t_{0}/\eta_{\rm C}\lesssim r_{0}\lesssim \eta_{\rm C}^{1/\mu}\lambda^{-1/\mu}$. This implies that the
asymptotic structure of the local map in the limit $t_{0}\to 0$,
therefore the main feature of the global map, is determined by only the
behavior of the null rays in the small but \textit{finite} region
$0<r\lesssim \eta_{\rm C}^{1/\mu}\lambda^{-1/\mu}$.
\section{Luminosity and energy of particle creation}
\label{Power}
We consider the massless scalar fields coupled to the scalar curvature $R$ as
\begin{eqnarray}
\left(\dalm-\xi R\right)\phi=0,
\label{scalar}
\end{eqnarray} 
where $\xi$ is an arbitrary real
constant. In particular,  the scalar fields with $\xi=0$ and $\xi=1/6$
are minimally and conformally coupled ones, respectively.
The stress-energy tensor of the scalar field in an 
asymptotically flat region is given by
\begin{eqnarray}
T_{\mu\nu}^{(\xi)}\simeq\nabla _{\mu}\phi\nabla _{\nu}\phi-\frac{1}{2}\eta_{\mu\nu}\nabla^{\alpha}\phi\nabla _{\alpha}\phi-\xi\nabla _{\mu}\nabla _{\nu}\phi^{2}+\xi \eta_{\mu\nu}\dalm\phi^{2}, \nonumber
\end{eqnarray}
where $\eta_{\mu\nu}$ is a flat metric.
To calculate the luminosity of emitted particles we need an estimate of the
vacuum expectation
value of the above stress-energy tensor. A suitable regularization is required in the calculation of the vacuum expectation value, because the stress-energy tensor is quadratic in fields at the same point.
The regularization for minimally and conformally coupled scalar fields is obtained in Ref.~\citen{FordParker} using the \textit{point-splitting regularization} scheme. This can easily be generalized to the above more general scalar fields to give
\begin{eqnarray}
L_{lm}^{(\xi)}(u)=\frac{1}{4\pi}\left[\left(\frac{1}{4}-\xi\right)\left(\frac{\mathcal{G}^{\prime\prime}}{\mathcal{G}^{\prime}}\right)^{2}+\left(\xi-\frac{1}{6}\right)\frac{\mathcal{G}^{\prime\prime\prime}}{\mathcal{G}^{\prime}}\right],
\label{luminosity}
\end{eqnarray}
where the prime denotes differentiation with respect to $u$, and $l$
($=0,1,\cdots$) and $m$ ($=0,\pm 1,\cdots,\pm l$) are angular momentum
and magnetic quantum numbers, respectively. The luminosity is the sum of
all these modes, but it diverges, because the
luminosity given in (\ref{luminosity}) is independent of ($l, m$). Such a
divergence is due to the fact that we ignored the 
back scattering by the potential barrier, which will reduce the emission for highly rotational modes. 
Hereafter, we omit the quantum numbers $(l,m)$, and it should simply be kept in mind that the above expression holds only for small $l$.
The total energy of emitted particles is estimated by integrating the luminosity with respect to $u$:
\begin{eqnarray}
E^{(\xi)}(u)\equiv\int^{u}_{-\infty}L^{(\xi)}(u^{\prime})du^{\prime}.\nonumber
\end{eqnarray}

\subsection{Quantum radiation in the non-self-similar LTB spacetimes: $0<\gamma<3$}
\label{sec:NSS}
As described in \S~\ref{sec:local-map}, we assume that the local map and the global map
have the same structure. This means that from Eq.~(\ref{eq:local-map}), the asymptotic form of the global map will take the form
\begin{eqnarray}
\mathcal{G}(u)\simeq v_{0}-A(u_{0}-u)-Ag\lambda^{3}(u_{0}-u)^{3\mu+1},\nonumber
\end{eqnarray}
where $u=u_{0}$ and $v=v_{0}$ are the CH and the ingoing null ray that
terminate at the NS, respectively, and $A$ and $g$ are constants. 
Keeping in mind that $\mu=(3-\gamma)/(3\gamma)$, one can calculate the luminosity as
\begin{multline}
L^{(\xi)}\simeq\left(\xi-\frac{1}{6}\right)\frac{3(3-\gamma)(3-2\gamma)}{4\pi\gamma^{3}}g\omega _{s}^{(3-\gamma)/\gamma}(u_{0}-u)^{-3+3/\gamma}\\
+\left[\xi-\frac{7\gamma-15}{36(\gamma-2)}\right]\frac{27(\gamma-2)(3-\gamma)}{4\pi\gamma^{4}}g^{2}\omega _{s}^{2(3-\gamma)/\gamma}(u_{0}-u)^{-4+6/\gamma}\\
+\textrm{O}\left((u_{0}-u)^{-5+9/\gamma}\right),
\label{eq:luminosity}
\end{multline}
where we have defined the ``frequency'' of the NSs as
\begin{eqnarray}
\omega_{s}\equiv\lambda^{1/\mu}=\lambda^{3\gamma/(3-\gamma)}.
\label{omegas}
\end{eqnarray}

In Appendix~\ref{frequency}, we show that the frequency~(\ref{omegas}) is
identical to that
defined in Ref.~\citen{HaradaIguchiNakao2000}. 
Depending on whether or not $\xi =1/6$, the leading-order term of the luminosity changes. Let us examine the time dependence of the luminosity and the total energy of emitted particles in detail for the cases $\xi\neq 1/6$ and $\xi=1/6$. 

In the case $\xi\neq 1/6$, the first term in
Eq.~(\ref{eq:luminosity}) dominates, except in the special case
$\gamma=3/2$. For $0<\gamma\leqslant1$, the leading term
vanishes as $u\to u_{0}$. 
For $1<\gamma<3$ and $\gamma\neq 3/2$, the luminosity diverges as the CH is
approached as a negative power of the time remaining to the CH. The
special case $\gamma=3/2$, in which the factor of the first term in
Eq.~(\ref{eq:luminosity}) vanishes, is divided into two cases, depending on whether $\xi\neq 1/4$ or $\xi=1/4$. If $\xi\neq 1/4$, the second term in Eq.~(\ref{eq:luminosity}) survives as a finite constant. If $\xi=1/4$, the factor of the second term also vanishes. Therefore, higher-order terms contribute to the luminosity, which is finite for $\gamma=3/2$. Therefore, in both cases that $\gamma=3/2$, the luminosity remains finite at the CH.
Now, let us examine the total energy of emitted particles.
For $3/2<\gamma<3$, the leading term is
\begin{eqnarray}
E^{(\xi)}\simeq-\left(\xi-\frac{1}{6}\right)\frac{3(3-\gamma)}{4\pi\gamma^{2}}g\omega_{s}^{(3-\gamma)/\gamma}(u_{0}-u)^{-2+3/\gamma},
\label{E1}
\end{eqnarray}
and therefore the total energy diverges as the CH is approached.
In particular, in the case of an analytic LTB solution ($\gamma=2$), the energy diverges as $(u_{0}-u)^{-1/2}$, which coincides with the result in Ref.~\citen{HaradaIguchiNakao2000}.
For $0<\gamma<3/2$, the total energy remains finite as the CH is approached.
In the special case $\gamma=3/2$, the energy also remains finite.
[See also Table~\ref{tb:minimal}(a).]

In the case of the conformally coupled scalar field, which is defined by $\xi=1/6$, the second term in
Eq.~(\ref{eq:luminosity}) dominates. It is found that in the case
$0<\gamma\leqslant 3/2$, the luminosity remains finite at most. While,
for $3/2<\gamma<3$, the luminosity diverges, due to the negative power of the time remaining to the CH. Let us examine the total energy of the emitted particles. For $2<\gamma<3$, we have
\begin{eqnarray}
E^{(1/6)}\simeq\frac{(3-\gamma)^{2}}{16\pi\gamma^{3}(\gamma-2)}g^{2}\omega_{s}^{2(3-\gamma)/\gamma}(u_{0}-u)^{-3+6/\gamma},
\label{E2}
\end{eqnarray}
which diverges as the CH is approached.
For $\gamma=2$, we have
\begin{eqnarray}
E^{(1/6)}\simeq\frac{3}{256\pi}g^{2}\omega_{s}\ln[\omega _{s}(u_{0}-u)]^{-1},
\label{E3}
\end{eqnarray}
and therefore the energy diverges logarithmically. This too is identical to the result in Ref.~\citen{HaradaIguchiNakao2000}. If $0<\gamma<2$, the energy remains finite, at most.
[See also Table~\ref{tb:minimal}(b).]

It follows that the radiation resulting from a conformally coupled scalar field is milder than that resulting from non-conformally coupled scalar fields for a given value of $\gamma$.
Such a consequence appears to result from the fact that the coupling of the conformal scalar field to gravity is weaker than that of the other scalar fields. 
In this relation, 
it should be pointed out that a conformally coupled scalar field must have non-zero and finite mass to be created in the early universe, while non-conformally coupled scalar particles are created regardless of their mass.\cite{BirrellDaviesFord}
\subsection{Quantum radiation in a self-similar LTB spacetime: $\gamma=3$}
\label{sec:SS}
The global map for self-similar LTB spacetimes ending in NS formation is calculated analytically in Ref.~\citen{BarveSinghVazWitten}. Subsequently, its main property was re-produced with the local-map method.\cite{TanakaSingh,MiyamotoHarada} \ The global map for null rays passing near the CH is given by
\begin{eqnarray}
\mathcal{G}(u)\simeq v_{0}-B(u_{0}-u)^{\alpha}\left[1+q(u_{0}-u)+\textrm{O}\left((u_{0}-u)^{2}\right)\right],\label{localmap-ss}
\end{eqnarray}
where $\alpha$, $B$, and $q$ are constants. The terms in the square brackets in Eq.~(\ref{localmap-ss}) together form an analytic function of $(u_{0}-u)$,\footnote{In previous works on particle creation during NS formation in the self-similar LTB spacetime, only the constant term in the square brackets in Eq.~(\ref{localmap-ss}) was considered.\cite{BarveSinghVazWitten,MiyamotoHarada,TanakaSingh} \ This was sufficient to obtain results. It is easy to calculate the higher-order terms using the local-map method and to show that they constitute an analytic function near the CH.
It is possible, however, that the emergence of the scale $q$ in Eq.~(\ref{localmap-ss}) from such a scale invariant spacetime as self-similar LTB solution indicates the breakdown of the local-map method. It is not clear to what extent the local-map method is valid.} \ The constant $\alpha$ depends only
on the parameter $\lambda$ in Eq.~(\ref{F}), and it has been shown to be greater
than unity for the region of $\lambda$ in which the singularity is naked.\cite{BarveSinghVazWitten} \ Using the global map~(\ref{localmap-ss}), we can compute the luminosity and energy of the particle creation as
\begin{eqnarray}
L^{(\xi)}&\simeq&\frac{(\alpha-1)(\alpha+1-12\xi)}{48\pi}(u_{0}-u)^{-2}+\frac{\alpha^{2}-1}{24\pi\alpha}q(u_{0}-u)^{-1},\label{luminosity-ss}\\
E^{(\xi)}&\simeq&\frac{(\alpha-1)(\alpha+1-12\xi)}{48\pi}(u_{0}-u)^{-1}+\frac{\alpha^{2}-1}{24\pi\alpha}q\ln q^{-1}(u_{0}-u)^{-1}.\label{energy-ss}
\end{eqnarray}
The first term in both Eqs.~(\ref{luminosity-ss}) and (\ref{energy-ss}) dominates, except in the special case that $\xi=(\alpha+1)/12$. Therefore, the luminosity and energy generically diverge as the inverse square and the inverse of the time remaining to the CH, respectively. However, in the special case $\xi=(\alpha+1)/12$, each second term in Eqs.~(\ref{luminosity-ss}) and (\ref{energy-ss}) dominates so that the power and energy diverge inversely and logarithmically, respectively.
This case, however, should be regarded as an accidental cancellation 
in the sense that the coupling constant $\xi$ happens to
be a special value $\alpha$, which is 
determined by the details of the collapse.
[See also Tables~\ref{tb:minimal}(a) and (b).]
\section{Discussion}
\label{Conclusion}
In this paper we have considered particle creation during the formation of shell-focusing NSs in a wide class of spherical dust collapse, which is described by the marginally bound LTB solutions. Each solution has a different initial density profile, and the resulting NSs have a variety of curvature strengths along the CHs. The luminosity and energy of particle creation have been estimated for each LTB solution and each scalar field that couples to scalar curvature in the linear form. 
The results are summarized in Tables~\ref{tb:minimal}(a) and (b). 

We first mention the validity of the approximations which have been made in this article. Next, we discuss the relations between the quantum radiation and the curvature strength of the NSs and also the manner in which the scalar fields couple to gravity. Finally, we discuss the implications of the present result to the CCH.

The analysis has been based on three assumptions: the validity of the
local-map method, the geometrical optics approximation, and quantum
field theory in curved spacetime.
The validity of each approximation seems to be open to debate.
(See Ref.~\citen{HaradaIguchiNakao2000} for discussion of the geometric optics approximation and Refs.~\citen{HaradaIguchiNakaoTanakaSinghVaz} and \citen{MiyamotoHarada} for quantum field theory in curved spacetime.) 
Here, we focus on the local-map method, on which our entire analysis is based.
The point is that the crucial factor of particle creation, the redshift of particles, must be determined by the geometry near the NS, while in Hawking radiation, the redshift is determined by the event horizon, which is a global object. It is unlikely that the global map has a structure that differs from that of the local map, since there is no singular feature in the map between the moments on the comoving observer at a finite distance and that of the null coordinates naturally defined at infinity. Indeed, in the models of the self-similar LTB,\cite{TanakaSingh,MiyamotoHarada} the analytic LTB~\cite{TanakaSingh} and the self-similar Vaidya~\cite{MiyamotoHarada}, the local-map method yields the correct results, which are obtained with the global map. 
For this reason, we have assumed the validity of the local-map method.

From the results, it is found that following statements hold for the generic naked-singular LTB spacetimes: \textit{the SCC along the CH is a sufficient condition for the
luminosity and energy of the created scalar particles to diverge as
the CH is approached; while, violation of the LFC is a sufficient 
condition for the luminosity and energy to be finite at the CH; if the NS does not satisfy the SCC but does satisfy the LFC, the luminosity and energy can be either divergent or finite.}
We only consider dust collapse for simplicity. However, the above statements regarding the curvature strength and the quantum radiation should be independent of the collapsing matter, because particle creation is a purely kinematic phenomenon and not directly related to the Einstein field equations. Therefore, we conjecture that the above statements hold for spherically symmetric collapsing spacetimes with \textit{any kind of collapsing matter}.
Of course, its validity should be verified or examined with known solutions ending in NS formation. The self-similar models which have been examined to this time, the collapse of a null-dust fluid,\cite{Vaidya,SinghVaz,MiyamotoHarada} a massless scalar field,\cite{Roberts,MiyamotoHarada} and a perfect fluid~\cite{OriPiran,MiyamotoHarada}, support this conjecture. There are many examples to be investigated: NS formation in counter-rotating particles,\cite{HaradaIguchiNakao1998,HaradaNakaoIguchi} non-self-similar null dust,\cite{Vaidya} null strange quark matter,\cite{HarkoChang} various matter fields in the critical collapse~\cite{Gundlach} and so on.
Here, we also mention the manner in which scalar fields couple to gravity. Although the quantum radiation due to the conformally coupled scalar field is less than that of other scalar fields, including the minimally coupled scalar field, the dependence of the amount of radiation on the manner of coupling is not so drastic as to modify the above statements.

Next, we consider the implications of our results for the CCH.
The diverging radiation from strong NSs corresponds to an instability of the strong NS formation. The system will enter into a phase where the backreaction from the quantum field to spacetime plays an important role.
While, the finite radiation from the weak NSs corresponds to a stability of the weak NS formation. This is striking, because the weak NSs seem to need another mechanism if they are to hide behind horizons. 
Of course, we cannot dismiss the possibility that the effect of backreaction suppresses the quantum radiation and that strong NSs appear, all things considered.

In the present analysis, we did not find a necessary and sufficient condition on the curvature
strength of NSs for the quantum radiation to be divergent or finite.
We believe that a new definition of the strength of (naked) singularities should be proposed from the viewpoint of the behavior of quantum fields on spacetimes rather than the viewpoint of the behavior of classical particles. 
Such a philosophy can also be seen in the wave-probe approach to NSs,\cite{WaveProbe} \ which is based on the theory of dynamics in non-globally hyperbolic spacetimes developed first by Wald.\cite{WaldIshibashi} \ With regard to this point, there is room for further investigation.

\section*{Acknowledgements}
We are grateful to Kei-ichi Maeda for his continuous encouragement. U.~M. thanks L.~H.~Ford for a helpful comment on the coupling of scalar fields and also thanks T.~P.~Singh and D.~A.~Konkowski for useful comments on the Cauchy horizon instability. Thanks are due to A.~Ishibashi for thoughtful comments on the implications and applications of the results. H.~M. was supported by a Grant for The 21st Century COE Program (Holistic Research and Education Center for Physics 
Self-Organization Systems) at Waseda University.
T.~H. was supported by JSPS.


\appendix
\section{Nakedness of the Singularity}
\label{nakedness}
In order to determine whether or not the singularity is naked, we investigate the future-directed outgoing null geodesics emanating from the singularity at $(t, r)=(0, 0)$. We find the asymptotic solutions that obey a power law near the center~\cite{JoshiDwivedi} in the form
\begin{eqnarray}
t\simeq X_{0}r^{p},\label{nullray0}
\end{eqnarray}
where $X_{0}>0$ and $p\geqslant 1$ are constants. The latter condition is due to the fact that the orbit of the shell-focusing singularity is $t=t_{s}(r)=r$. After some straightforward calculations, for $\mu>0$ we find the asymptotic solution
\begin{eqnarray}
t\simeq \frac{\lambda}{\mu+1}r^{\mu+1}.
\label{nullray}
\end{eqnarray}
With Eq.~(\ref{nullray}) and the fact that the apparent horizon, which is defined by $F=R$, behaves as $t=t_{ah}(r)=r-2F(r)/3\simeq r$ for $\mu>0$ near the center, the singularity is at least locally naked. In the case of the self-similar case ($\mu=0$), a similar treatment is possible, and the singularity is known to be naked for small values of $\lambda$.\cite{JoshiDwivedi} \ We consider the situation in which the collapsing dust ball is attached to an outer vacuum region at a comoving radius $r=[\mathrm{constant}$], within which the null ray~(\ref{nullray}) is outside the apparent horizon. Then, the singularity is globally naked, and the weak version of CCH is violated.
\section{Frequency of the Singularity}
\label{frequency}
It is helpful to compare the gauge used in this paper with
one used in much of the literature. This comparison shows that $\omega_{s}$ defined in \S~\ref{Power} coincides with the characteristic frequency of the singularity introduced in Ref.~\citen{HaradaIguchiNakao2000}, up to a numerical factor.
Let us denote the comoving coordinates by $(\tilde{t}, \tilde{r})$, in which $\tilde{r}$
 is chosen to coincide with the physical radius $R$ at the initial regular epoch $\tilde{t}=0$, i.e., we have $R(0, \tilde{r})=\tilde{r}$.
We assume that the mass function $F(\tilde{r})$ can be expanded near the regular center as
\begin{eqnarray*}
F(\tilde{r})=F_{1}\tilde{r}^{a}+F_{2}\tilde{r}^{b}+\cdots,
\end{eqnarray*}
where $a$ and $b$ are constants satisfying $a<b$. Then the initial density profile can be written
\begin{eqnarray}
\rho(0,\tilde{r})=\frac{a F_{1}}{8\pi}R^{a-3}+\frac{b F_{2}}{8\pi}R^{b-3}+\cdots.\label{Ftilde}
\end{eqnarray}
Comparing Eq.~(\ref{Ftilde}) with Eqs.~(\ref{expanded-rho}) and (\ref{coefficients}),
we obtain the powers and coefficients of $F(\tilde{r})$ as
\begin{eqnarray*}
a=3,&\;\;\;&b=\frac{3(3\mu+2)}{3\mu+1},\\
F_{1}=\frac{4}{9t_{\rm in}^{2}},&\;\;\;&F_{2}=-\frac{8(\mu+1)^{3/(3\mu+1)}}{9\lambda^{3/(3\mu+1)}(-t_{\rm in})^{(9\mu+5)/(3\mu+1)}}.
\end{eqnarray*}
It is found that the power $b$ satisfies $3<b\leqslant6$ for $\mu\geqslant 0$.
In Ref.~\citen{HaradaIguchiNakao2000}, Harada~\textit{et al.} determined the characteristic frequency of the naked singularity in the analytic model ($\mu=1/6$) using physical consideration. It is easy to repeat their computations for general values of $\mu>0$. One possible quantity, composed only of $F_{1}$ and $F_{2}$, which is independent of the choice of the initial time slice and has the dimension of frequency is 
\begin{eqnarray}
F_{1}^{(9\mu+5)/(6\mu)}(-F_{2})^{-(3\mu+1)/(3\mu)}&&\nonumber\\
&&\hspace{-3cm}=(\mu+1)^{-1/\mu}\left(\frac{2}{3}\right)^{(9\mu+5)/(3\mu)}\left(\frac{9}{8}\right)^{(3\mu+1)/(3\mu)}\omega_{s},\label{omegas2}
\end{eqnarray}
where we have used Eq.~(\ref{omegas}).
In terms of $\gamma$, the quantity (\ref{omegas2}) can be written as follows:
\begin{eqnarray}
\Omega_{\gamma}(\lambda)\equiv F_{1}^{(2\gamma+9)/(2(3-\gamma))}(-F_{2})^{-3/(3-\gamma)}.\nonumber
\end{eqnarray}
In the case of the analytic LTB model ($\gamma=2$), we have
\begin{eqnarray*}
\Omega_{2}(\lambda)=F_{1}^{13/2}(-F_{2})^{-3}\label{omega-analytic},\\
\end{eqnarray*}
which coincides with the frequency defined in Ref.~\citen{HaradaIguchiNakao2000}, up to a numerical factor. This shows that it is valid to define $\omega_{s}\equiv\lambda^{1/\mu}$ as the frequency of a singularity. In the self-similar LTB solution ($\mu=0$), such a quantity does not exist, because of the scale-invariant nature of self-similar spacetimes.

\begin{table}[htbp]
\begin{center}
\caption{The relations among the curvature strength of naked
 singularities, the luminosity, and the energy
 of scalar fields near the Cauchy horizons (a) for the
 non-conformally coupled scalar fields ($\xi\neq 1/6$) and (b) for
 the conformally coupled scalar field ($\xi=1/6$). The constant $\gamma$
 parameterizes the initial density profile of the dust fluid.
SCC implies LFC.
}
\label{tb:minimal}

\vspace{0.3cm}
(a)\\
\begin{tabular}{|c||c|c|c|c|c|c|c|c|c|} \hline\hline
$\gamma$ & 0 & &3/4& \hspace{1cm} & 1 &  & 3/2 & \hspace{1cm}  & 3 \\ \hline
Strength & -- & Weak & \multicolumn{6}{c|}{LFC} & SCC \\ \hline 
Luminosity & -- & \multicolumn{4}{c|}{Finite} & Divergent & \multicolumn{1}{c}{Finite} & \multicolumn{2}{|c|}{Divergent} \\ \hline
Energy & --& \multicolumn{6}{c}{Finite} & \multicolumn{2}{|c|}{Divergent} \\ \hline\hline
\end{tabular}

\vspace{0.5cm}

(b)\\
\begin{tabular}{|c||c|c|c|c|c|c|c|c|c|} \hline\hline
$\gamma$ & 0 & &3/4& \hspace{1cm} & 3/2 & \hspace{1cm} & 2 & \hspace{1cm}  & 3 \\ \hline
Strength & -- & Weak & \multicolumn{6}{c|}{LFC} & SCC \\ \hline 
Luminosity & -- & \multicolumn{4}{c|}{Finite} &
 \multicolumn{4}{c|}{Divergent} \\ \hline
Energy & --& \multicolumn{5}{c}{Finite} & \multicolumn{3}{|c|}{Divergent} \\ \hline\hline
\end{tabular}
\end{center}
\end{table}
\end{document}